# A POSSIBLE HYBRID COOLING CHANNEL FOR A NEUTRINO FACTORY*

J. C. Gallardo, BNL, Upton, NY 11973, U.S.A.
M. S. Zisman, LBNL, Berkeley, CA, 94720, U.S.A.

## Abstract

A Neutrino Factory requires an intense and well-cooled (in transverse phase space) muon beam. We discuss a hybrid approach for a linear 4D cooling channel consisting of high-pressure gas-filled RF cavities—potentially allowing high gradients without breakdown—and discrete LiH absorbers to provide the necessary energy loss that results in the required muon beam cooling. We report simulations of the channel performance and its comparison with the vacuum case; we also briefly discuss technical and safety issues associated with cavities filled with high-pressure hydrogen gas. Even with additional windows that might be needed for safety reasons, the channel performance is comparable to that of the original, all-vacuum Feasibility Study 2a channel on which our design is based. If tests demonstrate that the gas-filled RF cavities can operate effectively with an intense beam of ionizing particles passing through them, our approach would be an attractive way of avoiding possible breakdown problems with a vacuum RF channel.

## INTRODUCTION

Studies of the performance of high-gradient RF cavities in a strong axial magnetic field have been under way for several years [1] as part of the U.S. Neutrino Factory and Muon Collider Collaboration (NFMCC) R&D program. The motivation for such studies is that the cooling channel of a Neutrino Factory (or Muon Collider) requires this mode of operation.

Initial tests made use of an 805-MHz RF cavity operated in vacuum, with its beam irises terminated with Be windows. The Be windows result in an accelerating gradient that is nearly the same as the peak surface field in the cavity, thus improving the maximum accelerating gradient by roughly a factor of two compared with an open-iris cavity. While this cavity operated very well, providing 40 MV/m in the absence of magnetic field, its maximum gradient degraded in the presence of a strong magnetic field, as indicated in Fig. 1.

In contrast, a cavity filled with high-pressure hydrogen gas demonstrated [2] no such degradation, as shown in Fig. 2. With this in mind, we explore here the predicted performance of gas-filled cavities in a "hybrid" channel whose optics and energy loss material are the same as those of the Study 2a vacuum channel [3].

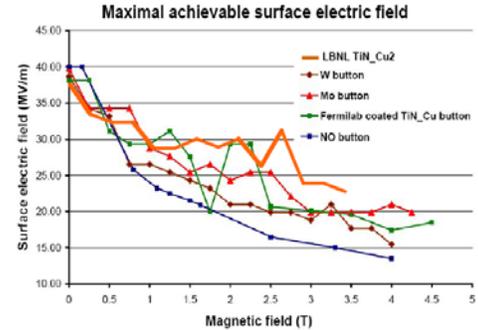

Figure 1: Maximum achievable surface field vs. applied axial magnetic field for a pillbox cavity with "button," showing a significant reduction compared with the no-field case.

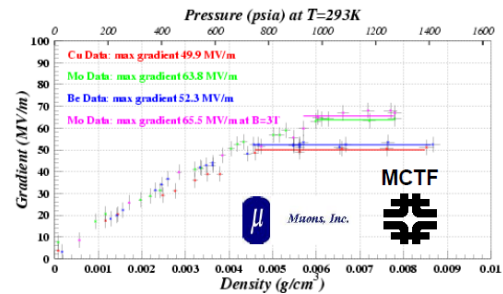

Figure 2: Achievable gradient in an $H_2$-gas-filled 805-MHz cavity vs. gas density. The upper lines show identical performance of molybdenum buttons with and without a 3 T applied magnetic field. See Ref. 2.

## IMPLEMENTATION STRATEGY

Because an RF cavity filled with high-pressure $H_2$ gas does not show any performance degradation in a magnetic field, this approach may be an attractive way to avoid such a problem. For the Study 2a channel, the required accelerating gradient is about 15 MV/m. If there is no frequency dependence of the behavior shown in Fig. 2, a pressure of ~10 atm should suffice to reach this gradient. To be conservative, in most of our calculations we assumed a gas pressure of 34 atm.

The absorber material in the Study 2a cooling channel is LiH. Each absorber is 1 cm thick. Our approach in the hybrid channel is to use the $H_2$ gas only for the purpose of permitting the cavity to operate at 15 MV/m. This amount of gas does not cause much energy loss, so LiH absorbers are still employed for that purpose. In the case of 34 atm gas, the energy loss from the gas contributes about 25% that of the LiH. To account for this, the thickness of all LiH absorbers in the channel was reduced by this amount, i.e., an absorber thickness of 7.5 mm was taken. For the

---
* Work supported by U.S. Dept. of Energy, Office of High Energy Physics, under Contract Nos. DE-AC02-05CH11231 (LBNL) and DE-AC02-98CH10886 (BNL).

10 atm case, the contribution to the energy loss is even less.

## PERFORMANCE COMPARISON

For our initial estimates, we compared the performance of the hybrid channel at 34 atm with that of the original Study 2a vacuum channel. The results, shown in Fig. 3, are quite encouraging. The emittance reduction of the hybrid channel is essentially the same as that of the Study 2a channel, and our transmission figure-of-merit, the number of muons per incident proton that fall within the acceptance of the downstream acceleration system, was slightly improved for the hybrid channel.

## USE OF ISOLATION WINDOWS

Given that a hybrid cooling channel might be 80–100 m in length, we have considered the possible need for intermediate isolation windows to separate the $H_2$ volumes into smaller segments. Engineering guidance [4] indicates that a tapered window is sufficiently strong without being inordinately thick, as illustrated in Fig. 4. To simplify our calculations, we implemented this shape in the ICOOL simulation code as a linear taper with a 35mm outer thickness and a 12 mm central thickness rather than a concave shape with 1-m radius, but this should not affect our results appreciably. Initially we considered Ti windows, but these substantially degraded the performance of the channel. We subsequently adopted Be windows, which appear relatively benign.

Figure 5 shows the performance of the hybrid channel with 17 Be windows, using the design concept shown in Fig. 4. We see that the emittance reduction is basically unaffected by the isolation windows. Transmission is also comparable to the vacuum case, except that the channel with Be windows reaches saturation more quickly than does the vacuum channel. If Ti windows were chosen instead, the performance is very poor (see Fig. 6); there is little emittance reduction and substantial particle loss.

We also compared the performance of the channel at a reduced gas pressure of 10 atm. If the thickness of the isolation windows is left unchanged, we find (see Fig. 7) that the preferred thickness of the LiH absorber is about 6.5 mm. Compared with the results in Fig. 5, the transmission has improved to be as good as that of the vacuum case.

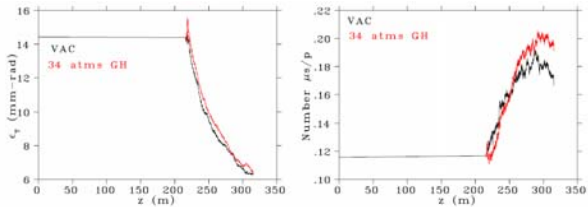

Figure 3: (left) Emittance reduction in hybrid channel (red line) and original Study 2a vacuum channel (black line); (right) number of muons per proton within downstream acceptance for both cases.

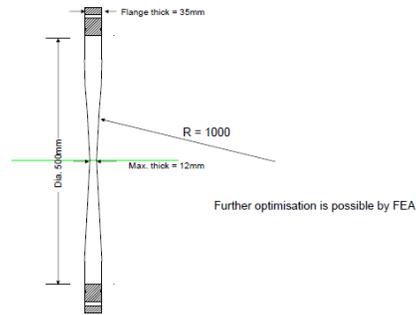

Figure 4: Schematic design for isolation window.

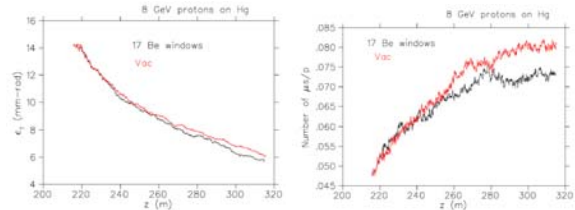

Figure 5: (left) Emittance reduction in hybrid channel with Be isolation windows (black line) and original Study 2a vacuum channel (red line); (right) number of muons per proton within downstream acceptance for both cases.

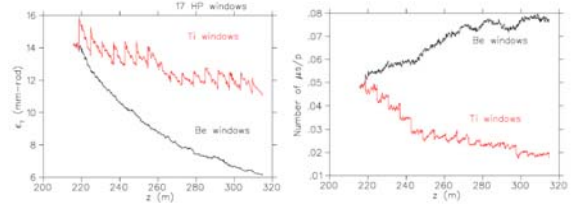

Figure 6: (left) Emittance reduction in hybrid channel with Be isolation windows (black line) or Ti windows (red line); (right) number of muons per proton within downstream acceptance for both cases.

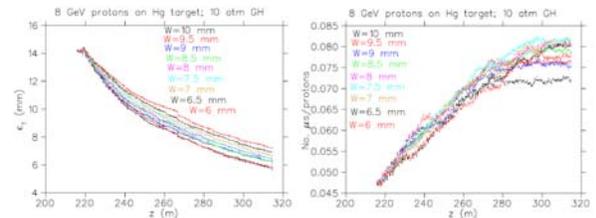

Figure 7: (left) Emittance reduction in hybrid channel with Be isolation windows for 10 atm of $H_2$ gas. The various curves represent different thicknesses assumed for the LiH absorbers; (right) number of muons per proton within downstream acceptance for various LiH thicknesses. Best performance results at ~7.5 mm thickness.

## IMPLEMENTATIONS

From the previous section we see that, if low-Z materials are acceptable, it is feasible to implement a modular hydrogen system having isolation windows and independent gas supplies. Materials issues are clearly important, most notably the question of hydrogen embrittlement of structural components. It is known that Al and Be-Cu are resistant to hydrogen embrittlement, but other cooling channel materials, such as Be, Cu, and LiH must be assessed.

Another unresolved issue concerns the operating temperature of the channel. There are several advantages to operating at liquid-nitrogen (LN) temperature. The required pressure for the same amount of insulating gas is reduced by a factor of nearly four, and the shunt impedance of the RF cavities will be roughly doubled. On the other hand, operating at LN temperature significantly complicates the engineering design of the channel. An insulating barrier is needed, heat must be removed from the RF cavities at a low temperature with a relatively poor coolant, and differential contraction of the various components will be a challenge.

In Fig. 8 we show an implementation of the hybrid cooling channel that provides a buffer vacuum. This version would be particularly attractive if the option of LN-temperature operation is considered important. In this scenario, the cavity itself must be designed as a pressure vessel. This is possible, of course, but there are many required penetrations (e.g., the power coupler and tuner) that must tolerate the high pressure. As indicated in Fig. 8, we assume that the LiH absorbers are fragile, and cannot be subjected to a large differential pressure. The gas filling and emptying schemes should ideally be designed to preclude such an eventuality by using common plumbing on both sides of the absorbers, as illustrated schematically in Fig. 8.

Figure 9 shows an alternative implementation in which the outer vessel serves as the pressure vessel. In this case, the RF cavity walls do not see any differential pressure, so the cavity can be designed more conventionally than that required for the implementation illustrated in Fig. 8. Here too, we assume the plumbing is designed to preclude developing any differential pressure during the filling and emptying processes. This version is relatively similar to the implementation being used in the MICE experiment [5], except that the outer vessel in the hybrid channel must be rated for high-pressure operation.

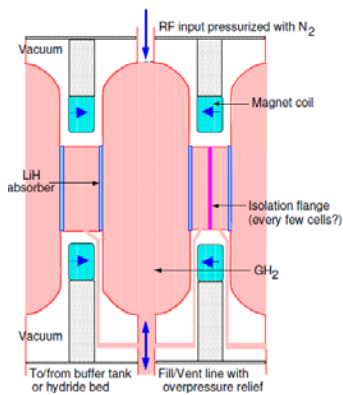

Figure 8: Schematic illustration of hybrid cooling channel utilizing a buffer vacuum system implementation.

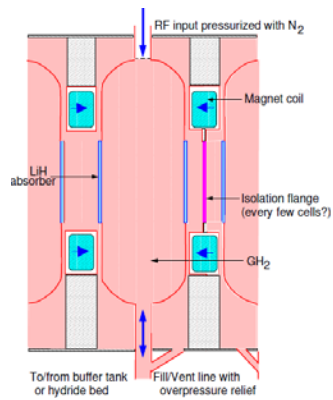

Figure 9: Schematic illustration of hybrid cooling channel implemented with a surrounding gas volume.

The implementation shown in Fig. 9 has the undesirable feature that the outside of the cooling channel would be classified as a hydrogen zone, with all of the complications that entails (e.g., explosion-proof electrical equipment requirements). In principle, the outer gas volume could be a relatively inert gas, such as nitrogen, which would help from a safety standpoint but makes it harder to ensure the absence of differential pressure during filling or emptying operations. No matter what the outer gas layer is, this implementation does not lend itself well to cryogenic operation.

## SUMMARY

We have looked at the implications of a hybrid linear cooling channel suitable for a Neutrino Factory or the initial stage of a Muon Collider, and find that it looks practical. Such a channel would eliminate the concern of degrading the RF cavity gradient due to the superimposed magnetic field, provided the gas-filled cavity operates properly when traversed by an intense beam of ionizing particles—something that remains to be demonstrated.